\documentclass[preprint]{aastex}

\received{}
\revised{}
\accepted{}
\journalid{}
\articleid{}
\paperid{}
\ccc{}
\begin{document}

\newcommand{\etal}{{\em et~al.\,}}
\newcommand{\f}{{RDCS1317+2911}}
\newcommand{\s}{{RDCS1350+6007}}
\newcommand\fdeg{\mbox{$.\!\!^{\circ }$}}

\title{Moderate Temperature Clusters of Galaxies from the {\tt RDCS}
and the High-Redshift Luminosity-Temperature Relation.
\altaffilmark{1,}\altaffilmark{2,}\altaffilmark{3}}

\altaffiltext{1}{Based in part on observations obtained at the W.M.\ Keck
Observatory}
\altaffiltext{2}{Based in part on observations obtained at Palomar
Observatory}
\altaffiltext{3}{Based in part on observations obtained with the
Chandra X-ray Observatory}

\author{B. P. Holden\altaffilmark{4,5}, S. A. Stanford\altaffilmark{4,5}}
\affil{Department of Physics, University of
California, Davis, CA 95616}
\email{bholden@igpp.ucllnl.org,adam@igpp.ucllnl.org}
\altaffiltext{4}{Participating Guest, Institute of Geophysics and
Planetary Physics, Lawrence Livermore National Laboratory}
\altaffiltext{5}{Visiting Astronomer, Kitt Peak National Observatory, 
National Optical Astronomy Observatory, which is operated by the
Association of Universities for Research in Astronomy, Inc. (AURA)
under cooperative agreement with the National Science Foundation.}

\author{G. K. Squires\altaffilmark{5}}
\affil{SIRTF Science Center, California Institute of Technology,
Pasadena, CA 91125}
\email{squires@ipac.caltech.edu}

\author{P. Rosati\altaffilmark{5}}
\affil{European Southern Observatory, Karl-Scharzschild-Strasse 2,
D-85748 Garching, Germany}
\email{prosati@eso.org}

\author{P. Tozzi}
\affil{Osservatorio Astronomico di Trieste, via G.B. Tiepolo 11,
I-34131, Trieste, Italy}
\email{tozzi@ts.astro.it}

\author{P. Eisenhardt}
\affil{Jet Propulsion Laboratory, California Institute of Technology,
MS 169-327, 4800 Oak Grove Drive, Pasadena, CA 91109}
\email{prme@kromos.jpl.nasa.gov}

\and 

\author{H. Spinrad}
\affil{Astronomy Department, University of California, Berkeley, CA 94720}
\email{spinrad@bigz.berkeley.edu}

\begin{abstract}

We present our discovery observations and analysis of \f , $z =
0.805$, and \s , $z= 0.804$, two clusters of galaxies identified
through X-ray emission in the \texttt{ROSAT} Deep Cluster Survey
(RDCS).  \f\ has an unusual morphology in our {\tt Chandra}
observations, with an asymmetric surface brightness profile and a bend
in the distribution of X-ray emission.  In contrast, \s\ appears to be
more like low-redshift clusters, with $\beta = 0.49 \pm 0.06$ and
$r_{core} = 165 \pm 5$ kpc ($\Omega_m = 0.3,\ \Omega_{\Lambda} = 0.7,\
H_o = 65\ {\rm km\ s^{-1}\ Mpc^{-1}}$), though it also has an
elliptical, slightly asymmetric surface brightness profile.  We find
a temperature of $3.7^{+1.5}_{-0.9}$ keV and a bolometric luminosity
of $8.2^{+1.7}_{-1.6}\times 10^{43}\ {\rm erg\ s^{-1}}$ for \f , and a
temperature of $4.9^{+1.3}_{-0.9}$ keV and a bolometric luminosity of
$4.1^{+0.5}_{-0.4}\times 10^{44}\ {\rm erg\ s^{-1}}$ for \s .  Our
weak lensing analysis of \s\ confirms the general shape of the inner
density profile but predicts twice the mass of the model based on
the X-ray profile.  There are two possibilities for this
discrepancy, either there is a significant amount of mass near the
redshift of the cluster that has not yet fallen into the potential well
and shock heated the gas, or, as we only see the
X-ray emission from the core of the cluster, our $\beta$ model fails to
describe the true shape of the underlying potential.

We combine the X-ray luminosities and temperatures for RDCS clusters
of galaxies with such measurements of other clusters at high-redshift
($z>0.7$) and fit the luminosity-temperature relation.  We find no
statistically significant evolution in the slope or zero-point of this
relation at $z_{median} = 0.83$.  This result is in agreement with
models of intracluster medium evolution with significant pre-heating
or high initial entropy values.  Quantifying the bolometric
luminosity-temperature relation as $L = L_6 (1+z)^A (T/6
keV)^{\alpha}$, we find $\alpha = 2.9 \pm 0.4$, $L_6 = 8.7 \pm 0.9
\times 10^{44}\ {\rm erg\ s^{-1}}$ and $A = 0.3 \pm 0.2$, or $A = 0.4
\pm 0.2$, depending on which low-redshift luminosity temperature
relation we compare with.  With this result, we rule out at the
5$\sigma$ level the self-similar scaling model of intracluster medium
evolution.  We discuss how low-temperature, high-redshift clusters of
galaxies will allow us to improve on this result and announce the
discovery of two such objects, CXOU J0910.1+5419 and CXOU J1316.9+2914.

\end{abstract}
\keywords{galaxies: clusters: general --- catalogs --- cosmology:
observations --- galaxies: clusters: individual (RDCS1350+6007,
RDCS1317+2911, CXOU J0910.1+5419, CXOU J1316.9+2914) --- X-rays }

\section{Introduction}

The luminosity-temperature relation for clusters of galaxies has
presented a puzzle for a number of years.  Using simple analytical
scaling relations, \citet{kaiser86} found that X-ray luminosity of
clusters of galaxies should scale as $L_x \propto T_x^2$.  However,
since the first compilations of the luminosities and temperatures for
clusters of galaxies \citep[for example]{es91}, it has been clear that
$L_x \propto T_x^{2.6 - 3.0}$, a steeper relation than predicted.

One way to break the scaling laws that predict $L_x \propto T_x^2$ is
to have non-gravitational energy injected into the intracluster medium
(ICM) before or during cluster formation.  This solution, called
pre-heating, was originally invoked to solve two related problems.
\citet{kaiser91} and \citet{evrard91} used pre-heating to explain the
apparent negative evolution of the X-ray cluster luminosity function
\citep{gioia90b,henry92} from the Einstein Medium Sensitivity Survey
in an $\Omega_m = 1$ Universe.  \citet{white91} also used pre-heating,
in the form of supernovae-driven galactic winds, to explain why groups
and low-mass clusters seem to have higher X-ray temperatures than
expected based on galaxy member velocity dispersions, or $1 <
\beta_{spec} \propto \frac{\sigma_v^{2}}{T_x} $.  This is the same
sort of behavior observed for the luminosity-temperature relation
where the temperature is too high for the luminosity when compared
with the highest temperature systems \citep{helsdon2000a}.  Recently
\citet{ponman1999} showed clearly that the entropy of the ICM in the
center of low-temperature clusters is greater than the value expected
from gravitational collapse.  Many authors have shown that this
additional entropy can reproduce many X-ray observational properties
\citep[for
example]{bower97,cavaliere97,cavaliere99,lloyddavies2000,lowenstein2000,wu2000,tozzi2001,bialek2001,borgani2001b,voit2001}.
However, there are many different physical processes that could break
the simple self-similar scaling, including heating from SN or from
AGN, or the removal of low-entropy gas via cooling \citep[for
example]{voit2001}.  To distinguish among these processes, the
observations of high-redshift groups and clusters to measure the
evolution in the observed scaling relations as a function of redshift
will prove crucial.

In order to tackle the thermal evolution of the ICM, we are examining
the highest redshift sample of X-ray selected clusters of galaxies
known, the \texttt{ROSAT} Deep Cluster Survey \citep[RDCS]{rosati98}.
Our goal is to observe all of the $z>0.8$ clusters in this catalog
with the Chandra \citep{axaf98} and XMM-Newton, thus creating a flux
limited sample at high redshifts with high quality ICM temperature and
luminosity measurements.  This will enable us to study both the
luminosity-temperature relation and to determine the distribution
functions of both of those physical measurements. In this paper we
discuss our results for two clusters of galaxies from the RDCS, \f\ in
\S 2 and \s in \S 3.  We then combine these data with our other
observations of $z > 0.8$ RDCS clusters of galaxies, along with a
number of other high-redshift clusters, and examine the
luminosity-temperature relation in \S 4.  In \S 5, we discuss the lack
of measured evolution of the luminosity-temperature relation and
discuss how future surveys can improve upon our understanding of the
thermal history of the ICM.  We use $H_0 = 65\ {\rm km\ s^{-1}\
Mpc^{-1}}$, $\Omega_{m} = 0.3$, and $\Omega_{\Lambda} = 0.7$.  At the
redshifts of \f\ and \s , this cosmology corresponds to approximately
8 kpc per second of arc.  Unless otherwise noted, errors represent
68\% confidence limits.

\section{RDCS1317+2911}

\subsection{Optical and Near-Infrared Data}

Our optical data for \f\ are a 1200 second $I$ band from January, 1996
using the Mayall 4m telescope with the Mosaic Imager, and an 1880
second Thuan-Gunn i band image using COSMIC \citep{kellscosmic98} on
the Palomar 5m telescope.  For the Palomar data, we had moderate
seeing, full-width at half-maximum $\simeq 1\farcs2$, in
non-photometric conditions.  The data covered a 9\farcm7 by 9\farcm7
field of view with 0\farcs 2856 pixels.  We observed this cluster in
$J$ and $K_s$ with the Prime Focus IR Camera \citep{jarrett94} with
the Palomar 5m in photometric conditions on 1998 March 24.  This
camera provides a $2\farcm1$ field of view with $0\farcs494$ pixels.
The flux scale was calibrated using observations of three UKIRT
standard stars obtained on the same night.  The data were taken using
a sequence of dither motions with a typical amplitude of 15\arcsec\,
and a dwell time between dithers of 30 seconds.  The data were
linearized using an empirically measured linearity curve, and reduced
using DIMSUM\footnote{ Deep Infrared Mosaicing Software, a package of
IRAF scripts available at ftp://iraf.noao.edu/contrib/dimsumV2}. In
Figure \ref{rxj1317s}, we show a color-composite image from these data
with the blue component from the $I$ band data, the green from the $J$
and the red from the $K_s$ data.

A catalog of objects in the $K_s$-band image was obtained using
SExtractor \citep{bertin96} after first geometrically transforming the
$K_s$ and $J$ images to match the $I$-band frame.  The resolution of
the IR images was also degraded slightly to match that of the $I$-band
image.  Objects were detected on the $K$-band image with the
requirement that that the object covers an area of 0.78 arcsec$^2$ and
must be 1.5 $\sigma$ above the background.  For reference the
3~$\sigma$ detection limit is $K \sim 21.3$ in the 2 arcsec aperture
used to measure colors.  All detected objects down to this limit were
inspected visually to eliminate false detections.  The catalog was
then applied to the $J$ and $I$ band images to obtain matched aperture
photometry.

\begin{figure}
\includegraphics[width=7.0in]{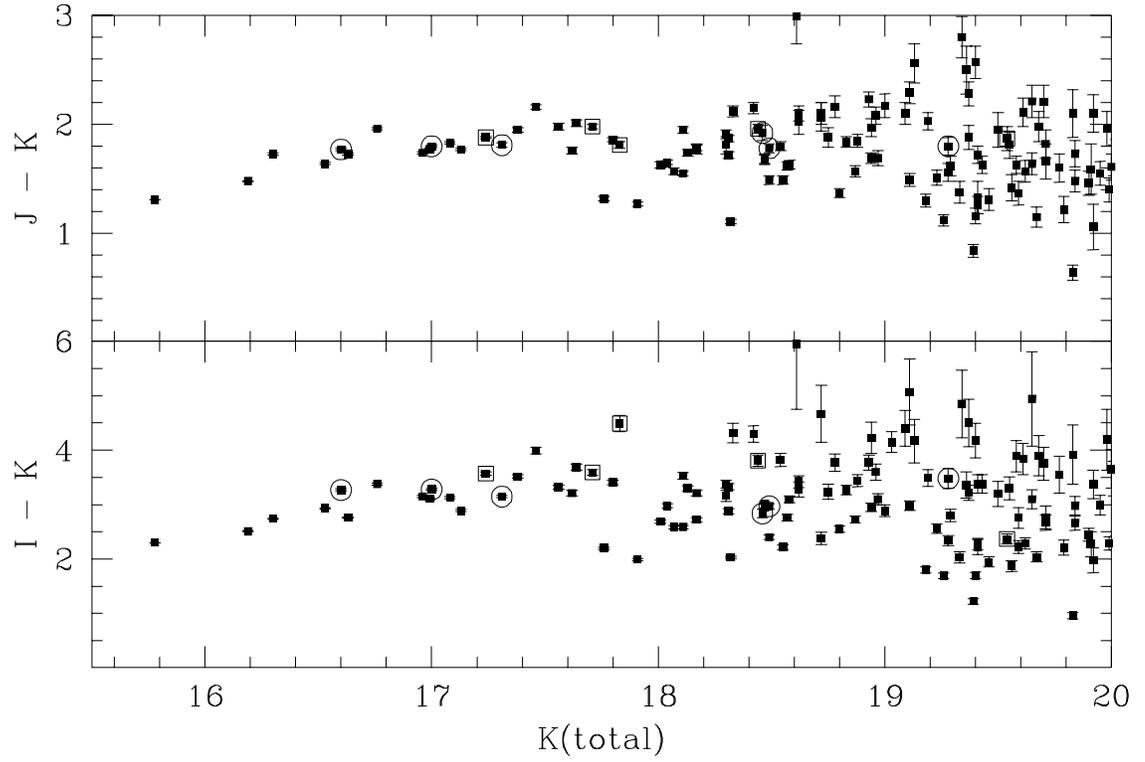}
\caption[holden.fig2.ps]{ 
The color-magnitude diagrams for \f.  
See \S 2.1 for details on how the photometry was constructed.  The
circled points represent spectroscopically confirmed cluster members
while the points with squares represent galaxies not at the redshift
of the cluster.
\label{rxj1317_cmd}}
\end{figure}

From these data we made color-magnitude diagrams, shown in Figure
\ref{rxj1317_cmd}.  In both colors, there is a clear red sequence of
galaxies, many of which are apparent in Figure \ref{rxj1317s}.  We
used the color-magnitude diagrams to choose targets for slit mask
observation.  \f\ was observed using the Low Resolution Imaging
Spectrograph \citep[LRIS]{oke95} on the Keck I telescope.  We obtained
the spectra using LRIS with a 150 line mm$^{-1}$ grating blazed at
7500 \AA , which has a dispersion of 4.8 \AA\ pixel$^{-1}$.  The mask
was observed in June of 1999 for three exposures of 1800 seconds with
small offsets along the slits between exposures.

We reduced the two-dimensional data using a set of IRAF\footnote {The
Image Reduction and Analysis Facility (IRAF) software is provided by
the National Optical Astronomy Observatories (NOAO), which is operated
by the Association of Universities of Research in Astronomy for
Research in Astronomy, Inc., under contract to the National Science
Foundation.} scripts optimized for LRIS observations.  These scripts
greatly reduce the fringing evident in LRIS spectra at long
wavelengths when used on dithered observations.  We extracted
one-dimensional spectra using the IRAF package APEXTRACT.  Individual
redshifts were measured by fitting the positions of lines and
the 4000\AA\ break.

\begin{figure}
\includegraphics[width=6.0in]{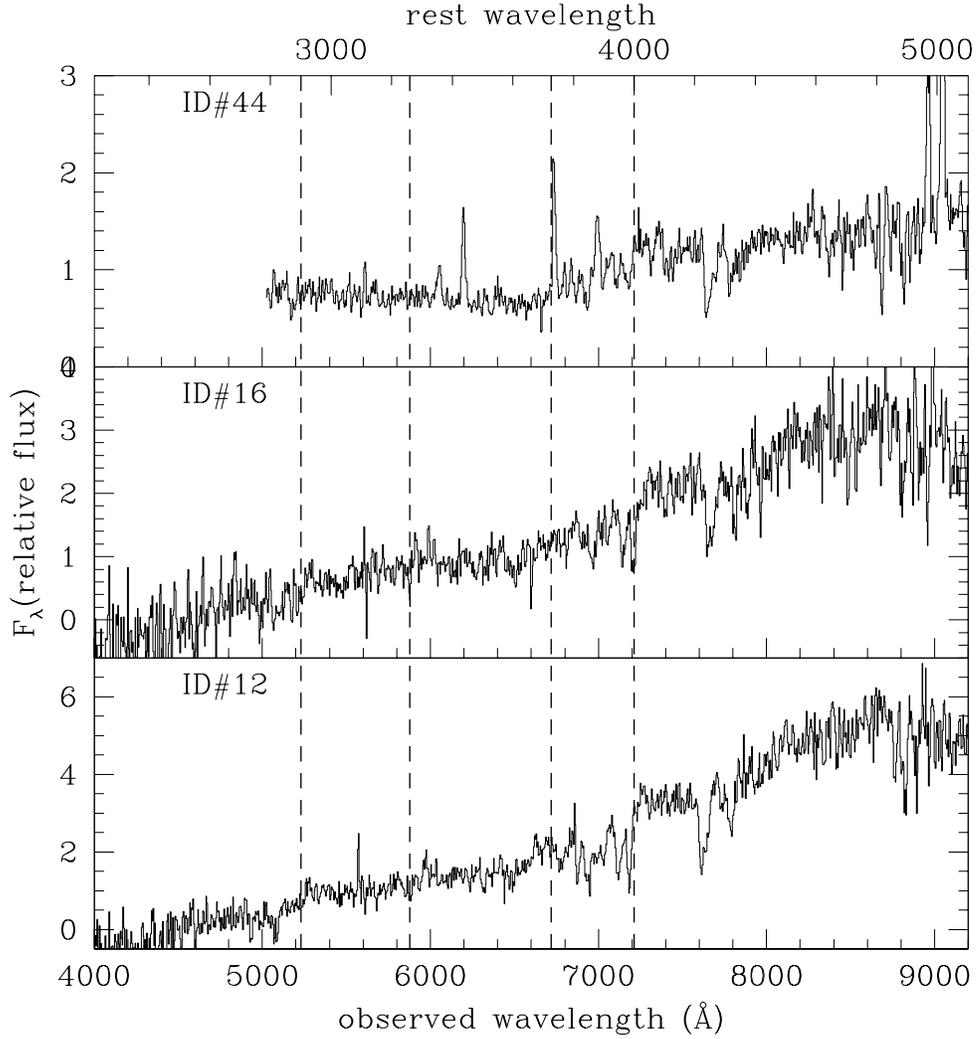}
\caption[holden.fig3.ps]{ We plot spectra for three of the members of
\f.  Each spectrum is shifted to the rest frame of the cluster.  The
four vertical dashed lines represent the break at 2900 \AA, the break
at 3260 \AA, \ion{O}{2} at 3727 \AA, and the break at 4000 \AA.  ID
\#44, at the top, shows evidence of \ion{O}{3} at 4959 \AA\ and 5007\AA .
\label{rxj1317_speca}}
\end{figure}

\begin{figure}
\includegraphics[width=6.0in]{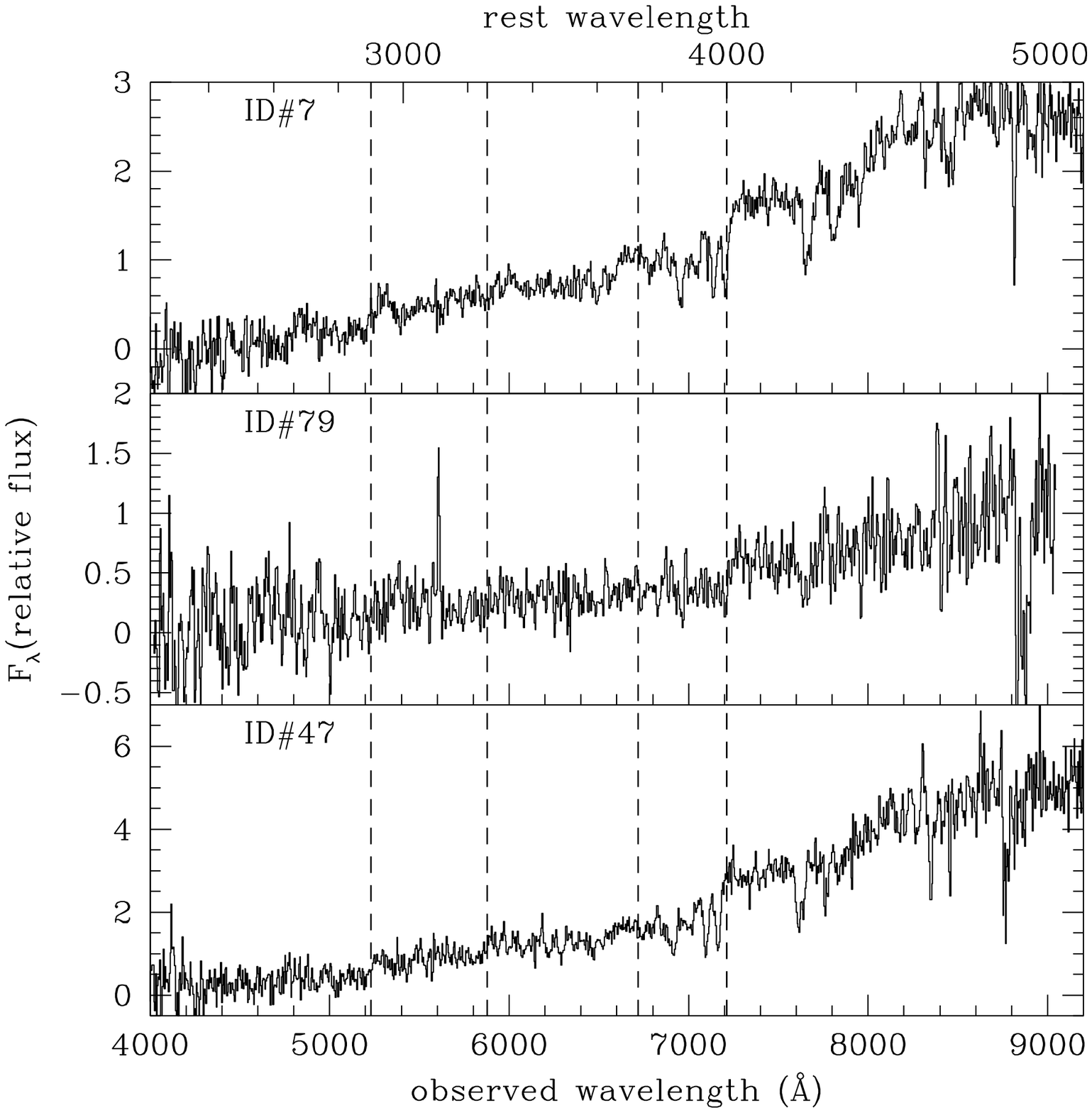}
\caption[holden.fig4.ps]{ 
See Figure \ref{rxj1317_speca}.
\label{rxj1317_specb}}
\end{figure}

We found six galaxies with $0.798 < z < 0.810$, which are circled in
Figure \ref{rxj1317_cmd}.  We show these spectra in Figures
\ref{rxj1317_speca} and \ref{rxj1317_specb}, plotted in the rest-frame
of the cluster.  We show, in addition, four vertical dashed lines
representing the break at 2900 \AA, the break at 3260 \AA, \ion{O}{2}
at 3727 \AA, and the break at 4000 \AA.  Using the biweight center
\citep{beers90}, we found $z = 0.805 \pm
0.002$ for the cluster.  We estimated the errors using the jackknife of
the biweight center as recommended in \citet{beers90}.

\subsection{X-ray Data}

\f\ was observed by {\tt Chandra}, using ACIS-I, as part of our
program to study the physical properties of the ICM in all $z>0.8$
RDCS clusters.  \f\ was observed for an effective exposure time of
111.3 ks (Obs ID 2228) on May 4th and 5th, 2001.  The observation was
done with the very faint mode when ACIS was operating at a temperature
of --120 C.

We processed the Level 1 data for each observation using the CIAO
v2.1.3 software. We cleaned the Level 1 event list for all events with
ASCA grades of 1, 5 and 7, filtered the data not in good time
intervals, removed bad offsets and removed bad columns. In addition,
we used the routines in clean55.1.0 \citep{vikhlinin2001} to remove
particle background events from data taken with the very faint mode.

We then removed, on a chip by chip basis, 3.3 second time intervals
when the count-rate exceeded three standard-deviations above the
average count-rate, $i.e.$ $\simeq$1.25 counts per second.  Once we
had created clean event files and corresponding exposure maps, we
merged the chip by chip event lists into one event file for the whole
pointing.  For the Level 1 processing, we used the ACIS calibration
files that were available on June 1, 2001.

\subsection{X-ray Imaging Results}

\begin{figure}
\vspace{6.0in}
\caption[holden.fig1.ps]{ A three-color montage of \f\ with the
0.5-2.0 keV X-ray data, smoothed with a 3\arcsec\ Gaussian,
overlaid as green contours.  The blue color is from a Gunn i band image,
with green and red from $J$ and $K_s$ band imaging.  The image is
2\farcm 5 by 2\farcm 5 on a side and the lowest contour level is 
$4\sigma$ above the background.  The image in the upper right shows
the 2.0-10.0 keV X-ray data smoothed with the same Gaussian.  The
hard point source seen in the inset is discussed in \S 2.3.  See
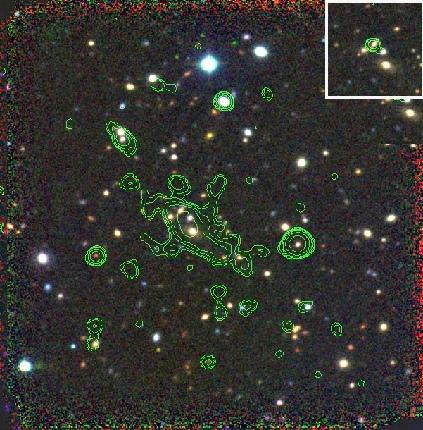.
\label{rxj1317s}}
\end{figure}

In Figure \ref{rxj1317s} we plot our 0.5-2.0 keV data for this
cluster, smoothed with a 3\arcsec\ Gaussian and overlaid on our ground
based imaging.  Overall, the cluster has a compact, highly elliptical
shape.  Using the Sherpa software package \citep{freeman2001}, we fit
an elliptical $\beta$ model to the 0.5-2.0 keV photons.  The resulting
model had a core radius of 0\farcs 1, and $\beta = 0.34$ with an
ellipticity $\epsilon = 0.53$ and is centered at $\alpha$ = 13${\rm
h}$17${\rm m}$21\fs 70 $\delta$ = +29\arcdeg 11\arcmin 18\farcs 1
(J2000).  The errors on all of these parameters are too small, $\sim$
1\% and the parameters appear quite different when compared with lower
redshift clusters of galaxies \citep{jones99}. This is because the
model does not describe the data well.  As seen in Figure
\ref{rxj1317s}, the X-ray emission from the cluster is not symmetric
but rather appears to have a strong kink.  Also, the peak in the X-ray
emission is off-center.  When we examine the difference between the
smoothed data and the best fitting model smoothed with the same
Gaussian, there is a clear over subtraction by the model in the center
and an under subtraction at the edges.

We measured an X-ray temperature for \f\ in the manner of our other
Chandra observations of RDCS clusters \citep{stanford2000,holden01c,
stanford2001}.  First we placed a circular aperture of 30\arcsec\ over
the center of the cluster as determined from the peak of the X-ray
emission.  We removed point sources from that aperture, including the
hard X-ray source shown in the inset of Figure \ref{rxj1317s}.  For a
background region, we extracted two rectangular apertures on either
side of the circular aperture.  For each background region, we also
removed point sources.  The two background regions were constructed to
have almost all of the flux land on the same node of the CCD as the
cluster.  For the background, we constructed weighted response matrix
files and auxiliary response matrix files using Alexey Vikhlinin's
calcarf/calcrmf tools (which have become part of the 2.2 version of
the CIAO package).  We fit the background regions with the {\tt
XSPEC} package of \citet{arnaud_xspec96} using a single component
power law not convolved with the auxiliary response matrix as a model
for the background from 0.5-6.0 keV with the addition of Gaussian at
2.1 keV to model the Au emission line.  Both regions were fit
simultaneously with only the normalizations allowed to vary.

We fit the data inside the 30\arcsec\ aperture using the Cash
statistic \citep{cash79} with a Raymond-Smith model \citep{raymond77}
and the above background model after rescaling for the relative areas.
There were a total of 313 events in the energy range of 0.5-6.0 keV.
We expected 137.7 events in the same energy range from the background.
The small number of events in the cluster, 175.3 net events, means we
cannot constrain the abundance or redshift from our data. We,
therefore, assumed a metal abundance of 0.3 solar \citep{horner2001}
and a redshift of $z=0.805$ for the cluster.  We included a galactic
absorption of $1.0\times 10^{20}\ {\rm cm^{2}}$ \citep{stark}.  We
found the best fitting temperature to be $3.7^{+1.5}_{-0.9}$ keV. We
found, inside the aperture, a flux of $6.5^{+0.6}_{-0.6}\times
10^{-15}\ {\rm ergs\ cm^{-2}\ s^{-1}}$ (0.5-2.0 keV) and
$1.1^{+0.1}_{-0.1} \times 10^{-14}\ {\rm ergs\ cm^{-2}\ s^{-1}}$
(0.5-6.0 keV).  We plot our best fitting spectrum in Figure
\ref{rxj1317_xray} with the background model subtracted and the data
grouped into bins of 10 events each.  Given the poor fit using the
$\beta$ model, we used a curve of growth from the 0.5-2.0 keV band to
determine the total flux.  Beyond a radius of 80\arcsec, the total
flux remains flat.  We find a total luminosity within 1 Mpc,
125\arcsec, of $3.0^{+0.6}_{-0.6}\times 10^{43}\ {\rm erg\ s^{-1}}$
(0.5 - 2.0 keV, in the rest frame of the cluster) and
$8.2^{+1.7}_{-1.6}\times 10^{43}\ {\rm erg\ s^{-1}}$ bolometric.  

Our errors are computed by 1000 Monte Carlo simulations using the
combined background and Raymond-Smith models.  For each simulation, we
use the XSPEC {\tt fakeit} command to make a simulated source with
background spectrum.  The same fitting process used for the real data
is then performed.  The errors are estimated from the distribution of
the resulting temperature, flux and luminosity values.  The median
temperature, flux and luminosity values agreed with the input values.

We also fit a power law model to the hard X-ray source in the core of
the cluster.  This source looks like a potential cluster member based
on the colors of the galaxy aligned with the X-ray emission but
currently it does not have a redshift.  As this hard source sits on
top of the cluster X-ray emission, we removed both the background and
the cluster emission when determining the spectrum.  We estimate a
flux of $1.1^{+0.2}_{-0.2}\times 10^{-15}\ {\rm ergs\ cm^{-2}\
s^{-1}}$ (0.5-2.0 keV) and $8.7^{+2.4}_{-2.1} \times 10^{-15}\ {\rm
ergs\ cm^{-2}\ s^{-1}}$ (2.0-10.0 keV) with a power-law slope of $0.6
\pm 0.3$ for the source.

\section{RDCS1350+60}

\subsection{Optical and Near-Infrared Data}

\begin{figure}
\vspace{6.0in}
\caption[holden.fig6.ps]{ \s\ with the 0.5-2.0 keV X-ray data,
smoothed with a 3\arcsec\ Gaussian, overlaid as green contours on top of an
three-color image.  The blue color is from an $R$ band image, with
green from an $J$ band image and the $K_s$ image for the red
component.  The field size is 2\farcm 3 by 2\farcm 3 with the lowest
contour at $5\sigma$ of the background. See 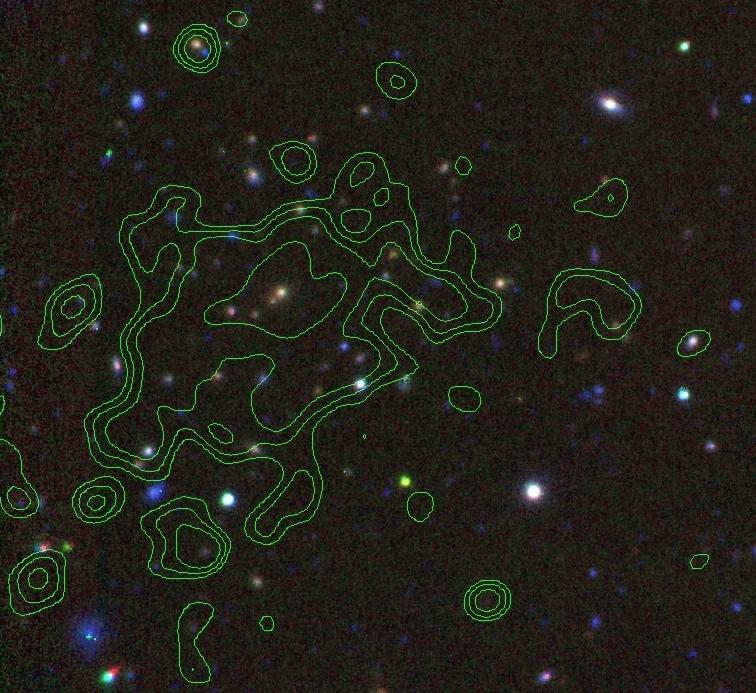.
\label{rxj1350s}}
\end{figure}

The optical counterpart for \s\ was originally observed using the
Mayall 4m Telescope with an $I$ band exposure.  Using that data, a
mask was made for the CryoCam spectrograph on the Mayall 4m Telescope
and three redshifts were measured in the cluster with $z=0.809$,
$z=0.808$ and $z=0.799$ from spectra taken in April, 1996.  Two masks
for LRIS were made via visual selection using the Kitt Peak $I$ band
image taken for the initial cluster discovery.  We observed each mask
twice with LRIS for 1200 seconds, with offsets between each
observation.  We took our spectra in June of 1999 using the 150 line
mm$^{-1}$ grating blazed at 7500 \AA.  We show an image of the cluster
in Figure \ref{rxj1350s}.

\begin{figure}
\includegraphics[height=6.0in]{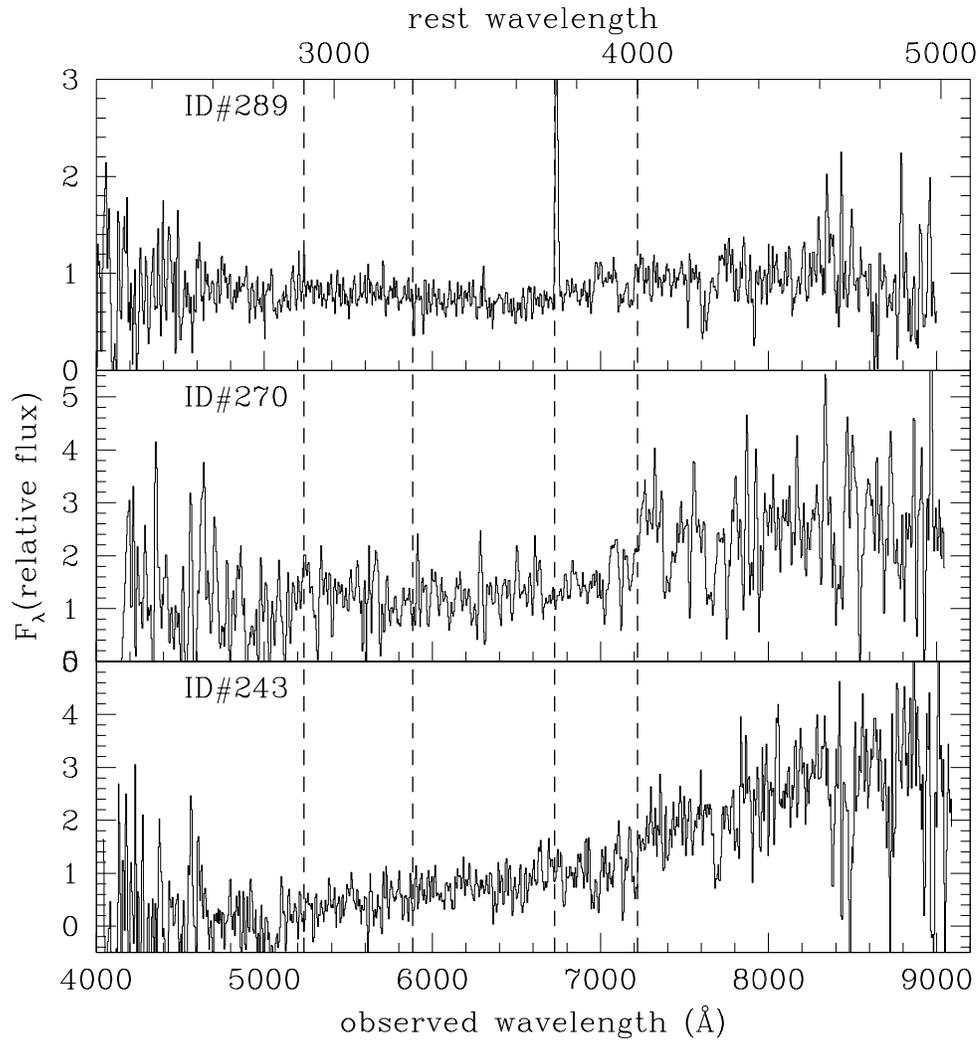}
\caption[holden.fig7.ps]{ We plot spectra for three of the members of
\s\ in the cluster rest-frame.  The vertical lines are the same as in Figure
\ref{rxj1317_speca}.
\label{rxj1350_speca}}
\end{figure}

\begin{figure}
\includegraphics[height=6.0in]{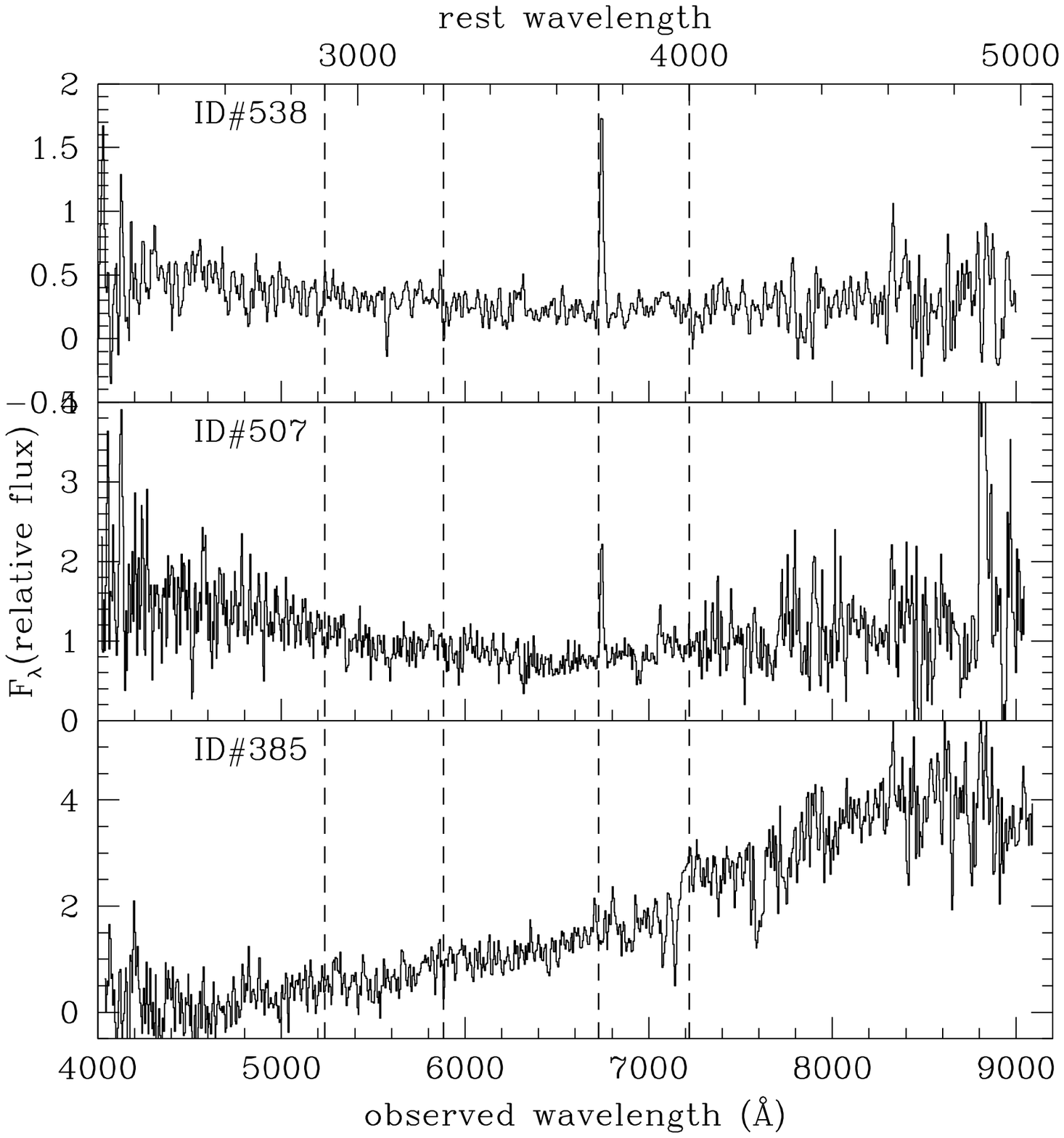}
\caption[holden.fig8.ps]{ 
See Figure \ref{rxj1350_speca}.
\label{rxj1350_specb}}
\end{figure}

The cluster redshift for \s\ was determined by searching for peaks in
the redshift distribution that consisted of galaxies close to the
X-ray centroid.  There are seven galaxies in the redshift range $0.800
< z < 0.809$ that are likely cluster members.  We used the same
procedure as we used for \f\ and find $z = 0.804 \pm 0.002$ for the
redshift of  \s .  We show those spectra in Figures
\ref{rxj1350_speca} and \ref{rxj1350_specb}.  Three galaxies in this
set of spectra were also those observed with CryoCam.  All three
objects had the same measured redshift in the LRIS spectra as was
originallly found in the Cryocam data.

\begin{figure}
\includegraphics[height=6.0in]{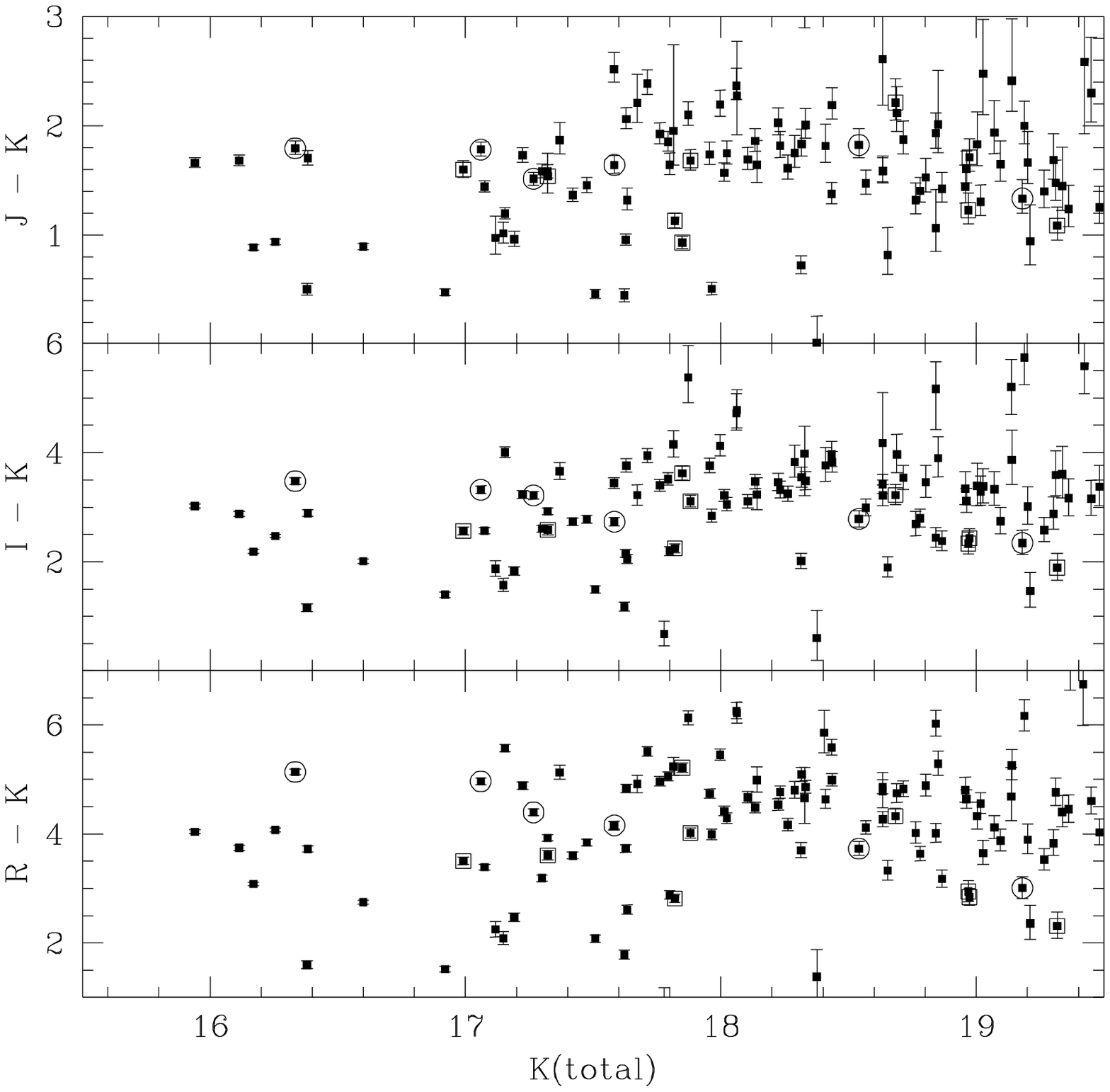}
\caption[holden.fig9.ps]{ The color-magnitude diagrams for \s.  See \S
2.2 for details on how the photometry was constructed.  The circled
points represent spectroscopically confirmed cluster members while the
points with squares are non-cluster members.
\label{rxj1350_cmd}}
\end{figure}

\s\ was observed June, 1999 for 9300 seconds with LRIS in the $R$ band
in 0\farcs 6 seeing.  These data were used as the blue color in the
image in Figure \ref{rxj1350s}.  Our $J$ and $K$ data, green and red
in Figure \ref{rxj1350s} respectively, were acquired with the
Simultaneous Quad IR Imaging Device \citep[SQIID]{ellissqiid92} on the
KPNO Mayall 4 meter telescope.  We observed \s\ for 4080 seconds in
June of 2001. The instrument has 0\farcs 39 pixels and, as its name
implies, takes $J$, $H$, $K$ and $L$ data simultaneously.  We use only
the $J$ and $K$ data in this paper.  The data were taken using a
sequence of dither motions with a typical amplitude of 15\arcsec\, and
a dwell time between dithers of 120 seconds.  The data were reduced
using DIMSUM.  Using SExtractor, objects were detected in the $K$-band
image with the requirement that the object covers an area of
0.8 arcsec$^2$ and must be 1.5 $\sigma$ above the background.  For
computing colors, a 3\farcs 3 aperture was used and a 5 $\sigma$
detection in that aperture is $K = 20.0$.  We plot the color-magnitude
diagrams for \s\ in Figure \ref{rxj1350_cmd}.  The circled objects are
galaxies with redshifts in the cluster.  Those points with squares
around are galaxies with redshifts that lie outside the cluster.

\subsection{X-ray Results}

\s\ was observed for an effective exposure time of 58 ks on August
29th and 30th, 2001 (Obs ID 2229) with ACIS-I instrument on the {\tt
Chandra} X-ray Observatory.  The observation and initial event
screening was done using the same parameters as \f\ as discussed in \S
2.2.

Glancing at Figure \ref{rxj1350s}, it is clear that \s\ has more
extended X-ray emission and has a much higher X-ray luminosity than \f
.  In Figure \ref{rxj1350s} we plot our 0.5-2.0 keV data for this
cluster, smoothed with a 3\arcsec\ Gaussian and overlaid on our ground
based imaging.  Once again, using the Sherpa package, we fit an
elliptical $\beta$ model to the 0.5-2.0 keV photons.  The resulting
model had a core radius of 20\farcs 3 $\pm$ 0\farcs 5 and $\beta =
0.49 \pm 0.06$ with an ellipticity $\epsilon = 0.39 \pm 0.03$ with an
X-ray centroid at $\alpha$ = 13${\rm h}$50${\rm m}$48\fs 55
$\delta$ = +60\arcdeg 07\arcmin 06\farcs 7 (J2000).  Other than the
high ellipticity, there is nothing unusual about these model
parameters.  The value for $\beta$ lies within the observed
distribution of \citet{jones99} but does create a problem.  At large
radii, a $\beta$ profile has the observed distribution of $\propto
(\frac{r}{r_c})^{-6\beta + 1}$, which, with our measured value of
$\beta$, yields a surface brightness profile $\propto
(\frac{r}{r_c})^{-2}$.  This means that the integrated flux of cluster
increases logarithmically with radius so we cannot use the $\beta$
model to calculate an aperture correction for a total luminosity at an
infinite radius.  Instead, we compute the total luminosity at a radius
of 1 Mpc as with \f.

For \s\ we measured the X-ray temperature and flux inside an ellipse
described by the best fitting $\beta$ model with the events taken from
within one core radius.  The one issue with this approach is that, as
shown in Figure \ref{rxj1350s}, the cluster is asymmetric with more
flux to the south-east than to the north-west of the cluster centroid.
As with \f , we extracted a background region and fit a model to the
background region. We chose a slightly more complicated model
consisting of a Gaussian and a broken power law\footnote{A broken
power law consists of two power law models with a shared
normalization, for three free parameters.  The two components meet at
a specific energy, the fourth parameter of the model.} not convolved
with the auxiliary response matrix and we fit the energy range of
0.8-6.0 keV.  We chose 0.8 keV for our cutoff in the energy range
instead of 0.5 keV because of a noticeable change in our background
regions below that energy.  To fit this change would require a much
more complicated background model than the broken power law and single
Gaussian we used.

\begin{figure}
\includegraphics[height=5.0in]{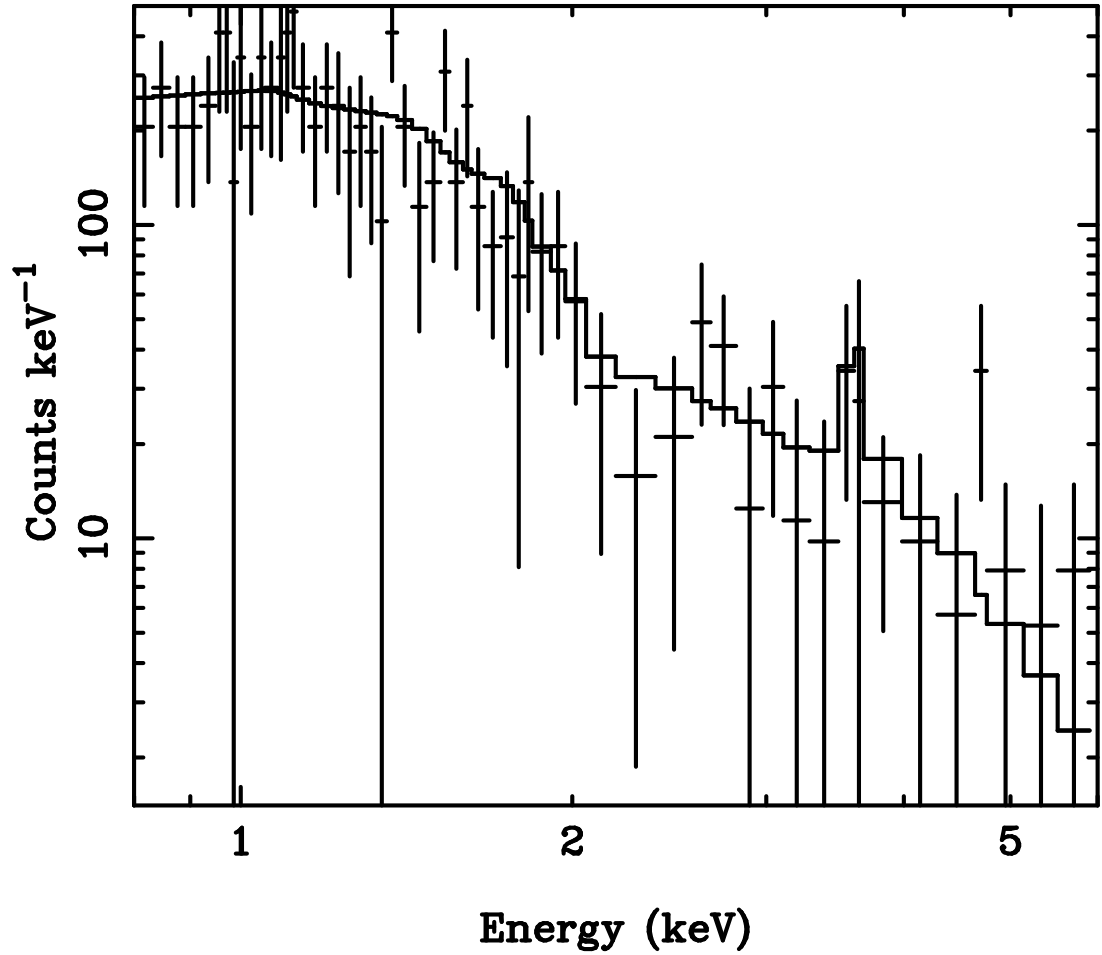}
\caption[holden.fig10.ps]{ 
Our best fitting spectrum for \s\ with
the background model subtracted and the data grouped into bins of five
events each. 
\label{rxj1350_xray}}
\end{figure}

We rescaled the normalization of the background model and fit the
cluster events with a Raymond-Smith model with a galactic absorption
of $1.8\times 10^{20}\ {\rm cm^{2}}$ \citep{stark}.  In this aperture
we found 354 events from 0.8 keV to 6.0 keV, the energy range we fit
the data and background over, while we expected 58 events from the
background in the aperture.  Freezing the redshift at $z = 0.804$ and
the metal abundance at 0.3 solar, we found the temperature to be
$4.9^{+1.3}_{-0.9}$ keV.  We plot our best fitting spectrum in Figure
\ref{rxj1350_xray} with the background model subtracted and the data
grouped into bins of five events each. The bump in the spectrum at
$\simeq$ 3.7 keV corresponds to the expected location of the 6.7 keV
Fe feature.  Inside the aperture, we find flux of $2.3^{+0.16}_{-0.16}
\times 10^{-14}\ {\rm ergs\ cm^{-2}\ s^{-1}}$ in the ROSAT hard band
(0.5 - 2.0 keV) and $4.1^{+0.38}_{-0.38} \times 10^{-14}\ {\rm ergs\
cm^{-2}\ s^{-1}}$ (0.5 - 6.0 keV).  The luminosity is
$1.4^{+0.2}_{-0.2} \times 10^{44}\ {\rm ergs\ s^{-1}}$ (0.5 - 2.0 keV,
in the rest frame of the cluster) and $4.1^{+0.5}_{-0.4} \times
10^{44}\ {\rm ergs\ s^{-1}}$ (bolometric) within a radius of 1 Mpc,
using the $\beta$ model to compute the flux inside that radius.  Our
errors are computed by 1000 Monte Carlo simulations, as was done for
\f , using the combined background and Raymond-Smith model.  For the
luminosity errors, we included our errors in the $\beta$ model fit to
the surface brightness distribution.

Because of the presence of the 6.7 keV Fe feature, we fit the data
assuming a free redshift and abundance.  We found a temperature of
$4.7^{+1.4}_{-1.0}$ keV, an abundance of $0.44^{+0.43}_{-0.24}$
relative to solar and $z = 0.84^{+0.03}_{-0.07}$ which is in good
agreement with our fixed values.  Choosing an ellipse of two core
radii, we found $4.1^{+0.7}_{-0.5}$ keV, an abundance of
$1.06^{+0.57}_{-0.37}$ relative to solar and $z =
0.87^{+0.01}_{-0.02}$.  This abundance is quite high but requires a
redshift in disagreement with the redshifts measured with optical
spectra.  This apparently high abundance comes from a number of events
with energies near the expected Fe feature.  These events are not
centrally concentrated like those events at the expected energy of the
Fe feature given the optically measured redshift of the cluster.
Therefore, we forced the redshift to agree with that of the optical
spectra.  That results in a temperature of $4.3^{+0.7}_{-0.5}$ keV and
an abundance of $0.58^{+0.40}_{-0.29}$.  For the rest of this paper we
will use our results with a fixed abundance and redshift within one
core radius, keeping the results from this cluster in agreement with
\f\ as well as those in \citet{stanford2000} and \citet{stanford2001}.
Nonetheless, we note that this cluster has a higher than expected
metal abundance.

\subsection{Weak lensing Analysis}

Using the 9300 second $R$ band Keck/LRIS exposure discussed in \S 2.2, we
modeled the mass distribution of \s\ based on the distribution of
image shears of distant field galaxies caused by weak lensing from the
cluster.  The lensing analysis followed standard techniques,
correcting for point spread function anisotropy following the Kaiser,
Squires \& Broadhurst method \citep{ksb95, hoekstra98}, and
calibrating losses due to seeing, pixelization, etc. as prescribed in
\citet{lk97}.

In order to compare the mass determined from the lensing with that
inferred via a standard X-ray analysis, we adopted the following
procedure.  Based on the X-ray data, we assumed an isothermal sphere
with a gas density profile given by a $\beta$ model fit.  In the X-ray
data, we lacked the signal to compare to a more sophisticated
elliptical model, so we employed only this circularly symmetric model.
The weak-lensing analysis determines the mass in concentric cylinders
along the line of sight. To compare with the X-ray inferred mass, we
re-fit a $\beta$ model, but fixed the ellipticity to zero.  The
resulting $\beta$ model has a core radius 15\farcs 6 $\pm$ 0\farcs 4,
and $\beta = 0.50 \pm 0.02$.  The core radius in the circularly
symmetric model is smaller because the symmetric model cannot fit the
light to the southeast of the core (see Figure \ref{rxj1350s}).

In the spherical model, the total mass enclosed inside a radius filled
with a gas in hydrostatic equilibrium, with density $\rho_g$ and
temperature $T$, is
\begin{equation} 
M_{\rm 3D}(r) = - \frac{k \, T(r)}{\mu m_p \, G} 
                  \left[ \frac{d\log \rho_g(r)}{d\log r} 
                   + \frac{d\log T(r)}{d\log r} \right] \, r 
\end{equation} 
where $r$ is the radius of interest, and $\mu$ is the mean molecular
weight, which we set to be $\mu = 0.59$.  An isothermal sphere with a
$\beta$ density profile yields the following mass distribution
\citep{fabricant84,henry93},
\begin{equation} 
 M_{\rm 3D}(r) = \frac{3 \beta k T r_{core}}{\mu m_p G}
                 \frac{(r/r_{core})^{3}}{1 + {(r/r_{core})^{2}}}
\label{eqn:m3dxray}
\end{equation}
where $M_{\rm 3D}(r)$ is the mass enclosed within a sphere of radius $r$.

However, a few subtleties arise when comparing the X-ray inferred mass
with that determined from the weak lensing analysis. First, the
lensing returns the mass along the line of sight compared with the
mass in a control aperture.  Hence, we need to calculate the
projected mass distribution from the parametric model fit to the X-ray
data. Second, mass determinations from weak gravitational lensing
analyses are necessarily lower estimates on the true mass along the
line of sight. The observed shear is unperturbed by the addition of
constant density sheets along the line of sight
\citep{gorenstein88}. Furthermore, the relatively small field of view
of these observations necessitates a differential determination of the
total mass. This is facilitated by the statistic
\citep{fahlman94,ksb95}
\begin{eqnarray}
\zeta_{WL}(\theta_1, \theta_2)& = & 2( 1 - \theta_1^2 / \theta_2^2 )^{-1}
                                  \int_{\theta_1}^{\theta_2} 
                                  d \ln(\theta) \langle \gamma_t \rangle \\
                            & = & \bar{\kappa}(\theta_1) - 
                                   \bar{\kappa}(\theta_1 < \theta <
                                   \theta_2) \nonumber  
\end{eqnarray}   
which measures the mean dimensionless surface density, $\kappa$,
interior to radius $\theta_1$, relative to the mean in an annulus
$\theta_1 < \theta < \theta_2$, and depends only on the measured
galaxy shear estimates, $\gamma_t$.

In order to do a meaningful comparison between the X-ray and lensing
mass determinations, we calculated an analogous estimator for the
X-ray model as follows: From equation \ref{eqn:m3dxray}, we find the
total mass density distribution \citep{fabricant84, henry93} to be
\begin{equation}
\rho_{tot} = \frac{3 \beta k T}{4 \pi \mu m_p G r_{core}^2}
\frac{3+(r/r_{core})^2}{[1 + (r/r_{core})^{2}]^{2}}.
\label{rho_tot}
\end{equation}
Hence, the projected mass density is given by
\begin{eqnarray}
\Sigma_X(b) & \equiv & \int_{-\infty}^{\infty} \rho_{tot}(b,z) \; dz \\
         & = &  \frac{ 3 \beta k T }{4 G \mu m_p r_{c}} \left( 1 +
         b^2/r_c^2 \right)^{-3/2} \times \left( 2 + \frac{b^2}{r_c^2}\right)
\end{eqnarray}
where $b$ is the distance from the cluster center in the plane of the
sky, and the z-axis defines the line-of-sight.  The equivalent X-ray
inferred mean (dimensionless) surface density interior to a given
impact parameter, $b$, relative to the mean in a control region $b
\leq b^\prime \leq b_{\rm max}$ is given by
\begin{equation} \zeta_X(b, b_max) = \frac{3 \beta k T}{2 \mu m_p G
r_c} \left( 1 - b^2/b_{\rm max}^2 \right)^{-1} \times \left[ ( 1 +
b^2/r_c^2)^{-1/2} - ( 1 + b_{\rm max}^2/r_c^2)^{-1/2} \right] /
\Sigma_{\rm crit}.  \label{eqn:zeta_xray} \end{equation} From either
the lensing or X-ray analysis, the projected mass at a given radius
$r$ is estimated as \begin{equation} M_{2D/WL,X}(r) = \Sigma_{\rm
crit} \times (\pi \, r^2) \times \zeta_{WL,X}(r, r_{\rm max}).
\label{eqn:m2d} \end{equation} The quantity $\Sigma_{\rm crit}$ is
defined by $\Sigma_{\rm crit}^{-1} = 4 \pi G \beta_l D_{l}$ where
$\beta = \langle D_{ls} / D_{s} \rangle$, and the distances $D_l,
D_{ls}, D_s$ are the angular diameter distance to the lens, between
the lens and the background source, and to the source,
respectively. For this analysis, we use the photometric redshift
distributions \citep{gwyn96,gwyn99,soto99} inferred for galaxies in
the Hubble Deep Fields to estimate $\Sigma_{\rm crit}$ for the
magnitude cuts selected for the galaxies in the lensing catalog.

\begin{figure}
\includegraphics[height=6.0in]{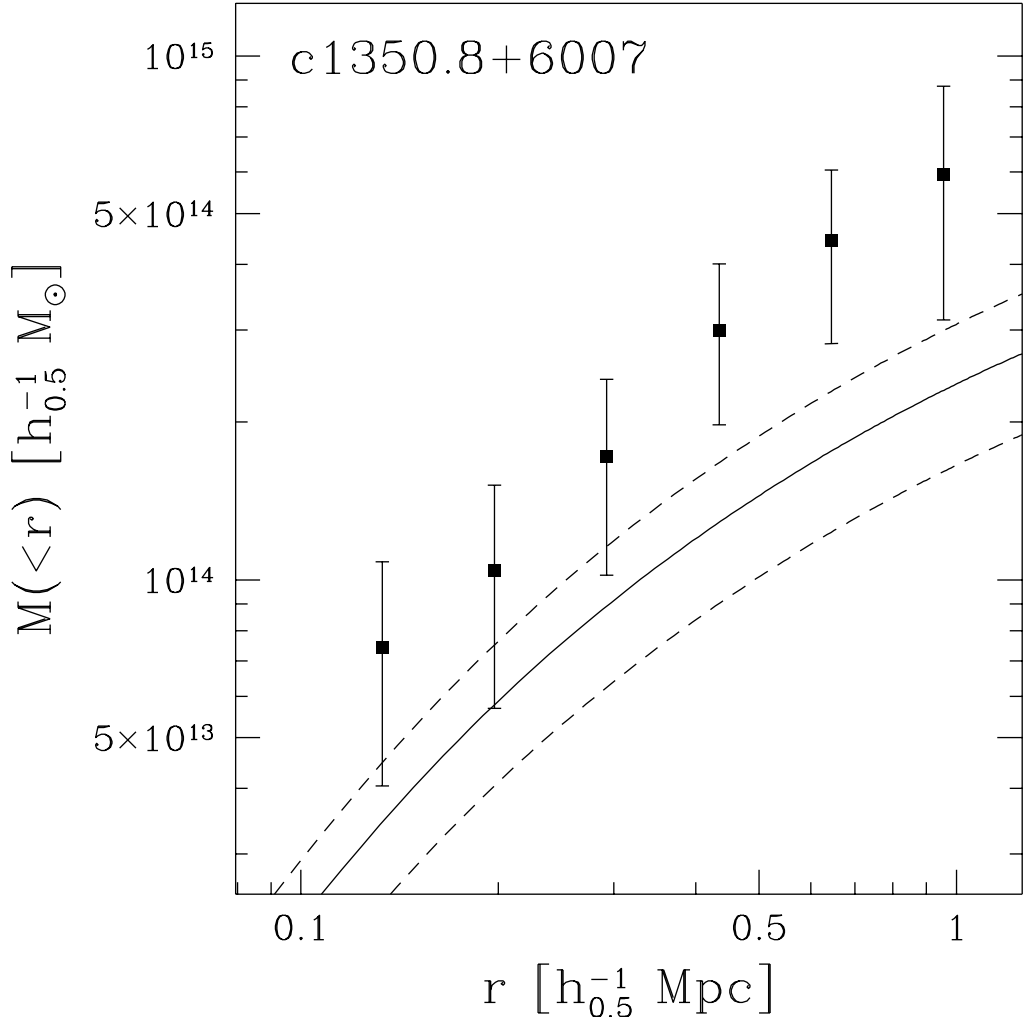}
\caption[holden.fig11.ps]{The total enclosed mass as a function of
projected distance from the cluster center.  The solid line represents
the enclosed mass based on assuming an isothermal gas following the
$\beta$ model distribution and the dashed lines represent the 68\%
confidence limits for that model.  The points are the enclosed mass
measurements from the weak lensing analysis.  Each weak lensing data
point represents the cumulative mass inside that aperture, so the
errors are correlated between the data points.
\label{weaklens}}
\end{figure}

We plot in Figure \ref{weaklens} the resulting radial, projected total
mass distributions calculated from equation \ref{eqn:m2d}, in
concentric cylinders along the line of sight. The points are the
results from the lensing, while the lines show the X-ray model,
adopting $\theta_c = 15\farcs6$, $\beta = 0.49$, $k\, T = 4.9 \; {\rm
keV}$, and $\mu = 0.59$.  We plot the 68\% confidence lines on either
side of the solid line representing the results from the best fitting
$\beta$ model.  The statistical uncertainty in the X-ray mass
determinations was calculated propagating the errors in quadrature.

It is apparent that our mass model based on the X-ray profile does not
predict as high a mass as does the weak lensing shear signal.  At
large radii, this result is expected, as we show in Figure
\ref{rxj1350s}. We have a non-azimuthally symmetric X-ray gas
distribution.  Therefore, our circularly symmetric X-ray model will
underestimate the total mass by neglecting the X-ray emission to the
southeast.  However, even at the core, our methods do not agree.  Weak
lensing depends on the projected mass along the line of sight, so we
could be observing a system with a significant mass in the foreground
that cannot seen in the X-ray emission.  One such example of this is
the cluster ZwCL 0024+17 where the gravitational arcs
\citep{tyson1998} and weak lensing \citep{smail1997} imply a large
mass near the cluster.  However, the X-ray analysis of
\citet{bohringer2000} find a small core radius, $r_{core} = 51\ {\rm
kpc\ ( H_o = 65\ km\ s^{-1}\ Mpc^{-1}})$ and $\beta = (0.43 - 0.55)$
(68\% confidence limits).  Such small values of $\beta$ and $r_{core}$
imply a small binding mass.  With the galaxy redshift measurements
from \citet{czoske2001}, it is clear that ZwCL 0024+17 is a merger
along the line of sight, thus explaining the large lensing mass but
the small gas mass implied by hydrostatic equilibrium.  Such a line of
sight merger could explain our small value for $\beta$.  A merger
could ``puff-up'' the gas, causing a smaller value for $\beta$ than
required by hydrostatic equilibrium.  Interpreting our results shows
the importance of using multiple methods of studying a cluster to
disentangle the underlying mass distribution such as in
\citet{zaroubi98}, \citet{holden01c} or \citet{zaroubi2001}.

\section{The $z \simeq 0.8$ Luminosity-Temperature Relation}

\begin{figure}
\includegraphics[height=6.0in]{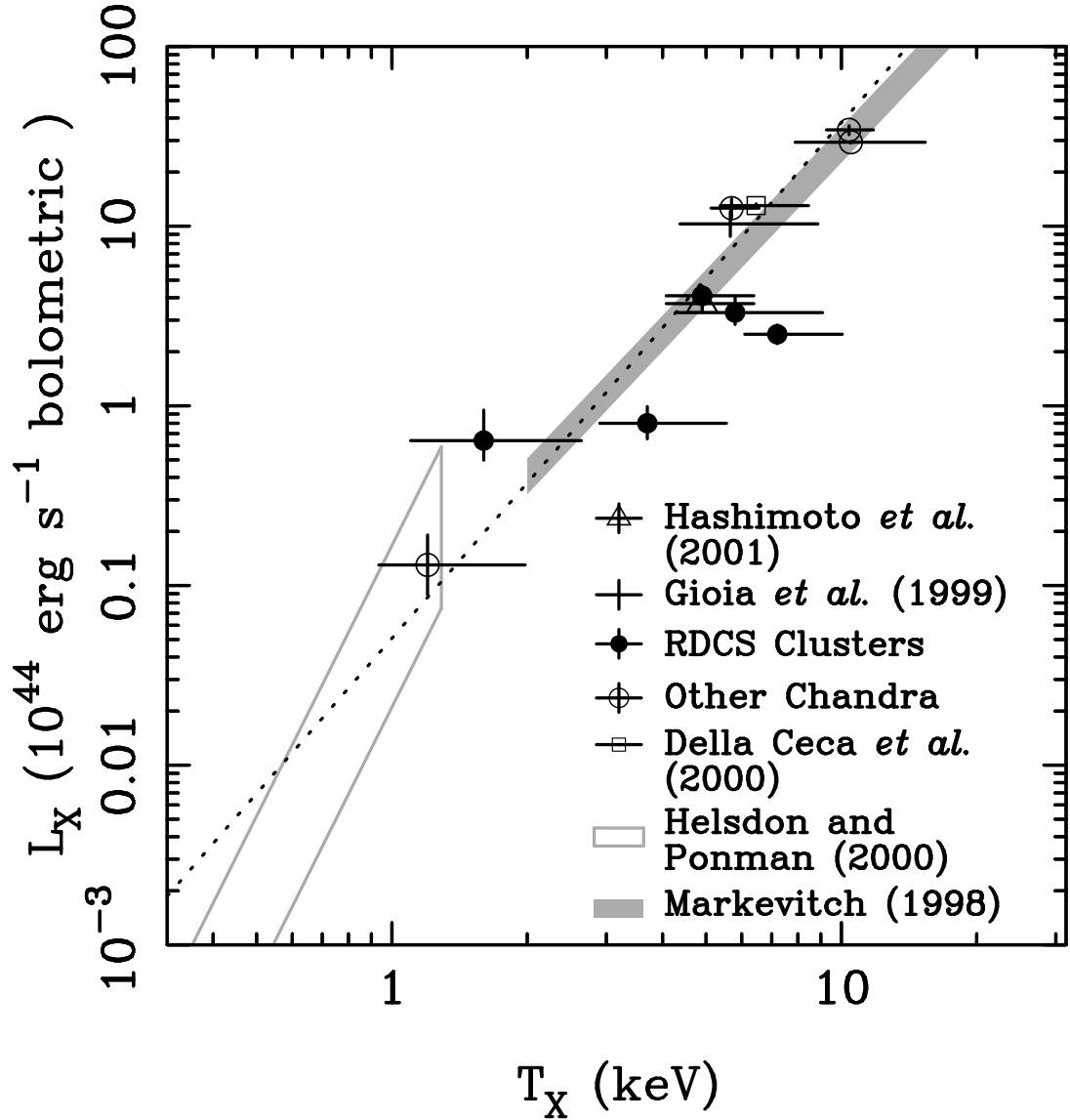}
\caption[holden.fig12.ps]{The luminosity-temperature relation
discussed in this paper.  Low-redshift group data, represented by the
outlined area, are from \citet{helsdon2000a}.  The low-redshift cluster
results, from \citet{markevitch98}, are plotted as the shaded region.
All other clusters are at $z>0.7$ and the dotted line represents the
fit to that sample given in the text.
\label{lt}}
\end{figure}

When we combined the clusters from \citet{stanford2000,stanford2001}
with those in this paper, we have a total of five $z>0.8$ clusters of
galaxies with {\tt Chandra} observations to date.  We will use this
RDCS sample in conjunction with other cluster measurements from the
literature, to estimate the luminosity-temperature relation at $z
\simeq 0.8$.  All clusters are listed in Table \ref{cl_sum}.  In
Figure \ref{lt}, we plot the luminosities and temperatures of the
above clusters.  The RDCS clusters are labeled with solid dots, RDCS
J0152.7-1357 by an open square, RX J1053.7+5753 by a diamond, and all
of the other clusters are represented by just error bars.  RX
J0152.7-1357 was observed using Beppo-SAX \citep{dellaCeca00}, RX
J1716+6708 with {\tt ROSAT} and ASCA \citep{gioia1999}, and RX
J1053.7+5753 was observed with XMM-Newton \citep{hashimoto2002}, so
our plot uses different symbols for those clusters.  All of the other
clusters were observed with {\tt Chandra}.  We also show the
low-redshift group data from \citet{helsdon2000a} and the low-redshift
cluster luminosity-temperature relation from \citet{markevitch98}.

For low-redshift clusters, there is a fair amount of agreement in the
slope of the luminosity temperature relation. \citet{allen98} estimate
the luminosity-temperature relation for a sample analyzed in a manner
similar to ours, a single temperature, isothermal model for the
cluster, which they refer to as Model A.  The sample spans $0.0258 \le
z \le 0.451$, with $z_{med} = 0.2045$.  Parameterizing the relation as
$T_x = P L_{x,bol}^{Q}$, \citet{allen98} find $Q = 0.325 \pm 0.061$,
using the bivariate correlated errors and intrinsic scatter (BCES)
algorithm from \citet{akritas96} for fitting models to quantities with
errors in both the x and y values.  Conventionally, the
luminosity-temperature relation is described by $L_{x,bol}\ ({\rm erg\
s^{-1}}) = L_6 (T_{x}/6\ {\rm kev})^{\alpha}$ where $L_6$ is the
luminosity of a 6 keV cluster. In this form, the best fitting slope
from \citet{allen98} is $\alpha = 3.1 \pm 0.6$.  \citet{markevitch98}
finds $\alpha = 2.64 \pm 0.16$ using a sample of $0.04 \le z \le
0.09$, with $z_{med} = 0.055$, clusters of galaxies that have been
corrected for cooling flows.  \citet{arnaud1999} find a similar
result, $\alpha = 2.88 \pm 0.15$ for a sample constructed assuming
isothermal temperatures but with strong cooling flow clusters removed.
This sample spans $0.0038 \le z \le 0.370$ with $z_{med} = 0.0522$.

Our sample is rather heterogeneous, containing a number of clusters
that appear to have multiple components (MS 1054.4-0321, RX
J1053.7+5753, \& RDCS0152.7-1357).  In addition, we have made no
corrections for cooling flow clusters.  This means that the results of
\citet{allen98} are the most comparable.  Using the data plotted in
Figure \ref{lt}, we fit the luminosity as a function of temperature
over the temperature range of 1 to 10 keV and found $\alpha = 3.0 \pm
0.4$ and $L_6 = 8.7 \pm 0.9 \times 10^{44}\ {\rm erg\ s^{-1}}$ using
the BCES method. This sample has a median redshift of $z_{med} =
0.83$.  If we trim the two clusters at $0.7 < z < 0.8$, we find a
steeper slope with $\alpha = 3.1 \pm 1.1$ and $L_6 = 8.2 \pm 1.0
\times 10^{44}\ {\rm erg\ s^{-1}}$ with $z_{med} = 0.86$.  Given the
large errors, both fits are compatible at the $1\sigma$ level with all
of the low-redshift estimates.

Assuming no evolution in the slope, we can instead look for a change
in the value for $L_6$.  The normalization of the
luminosity-temperature relation of \citet{arnaud1999} is $L_6 = 6.8
\pm 0.5 \times 10^{44}\ {\rm erg\ s^{-1}}$ while \citet{markevitch98}
finds $L_6 = 7.4 \pm 0.6 \times 10^{44}\ {\rm erg\ s^{-1}}$ (converted
to $H_o = 65\ {\rm km\ s^{-1}\ Mpc^{-1}}$).  Both normalizations are
lower than ours, though not by a statistically significant amount.
If we lower our slope to be the same as \citet{arnaud1999} or
\citet{markevitch98}, our values for the normalization of the
luminosity-temperature relation are roughly two standard deviations
higher.  

\section{Discussion}

In \citet{borgani2001}, the authors parameterize the evolution of the
luminosity-temperature relation with $L_{x,bol}\ {\rm erg\ s^{-1}} =
L_6\ (T_{x}/6\ {\rm kev})^{\alpha}\ (1+z)^A$ where $A$ is predicted to
be $-0.7 < A < 0.7$ in ICM models with additional entropy from
non-gravitational sources of heating \citep{tozzi2001} while $A = 1.5$
in self-similar models.  We find $A = 0.41 \pm 0.21$, when comparing
our whole $z> 0.7$ sample with \citet{arnaud1999} and, similarly $A =
0.27 \pm 0.22$ when comparing with \citet{markevitch98}.  Though our
results cannot accurately pin down the value of $A$, we can rule out
at the 5$\sigma$ level the evolution predicted for self-similar ICM
model.  Both of the values for $A$ we have above are lower than the
value of $0.60 \pm 0.38$ found by \citet{fairley2000}, though not by
more than one standard deviation.  

\citet{mushotzky97} constructed two samples, a $z < 0.1$ sample and a
$z > 0.14$ sample with a characteristic redshift of $z \simeq 0.3$.
\citet{mushotzky97} find no observed evolution with $A = 0.0 \pm 0.7$.
The agreement in the observed lack of evolution between a variety of
different samples observed with different instruments, solidifies the
evidence that the self-similar scaling model cannot explain ICM
formation.  Moreover, a value of A close to zero implies little
evolution of the cluster temperature (or mass) function based on the
observed mild evolution of the X-ray luminosity function. As discussed
in \citet{borgani2001}, this result strongly points toward low values
of $\Omega_m$.

Two of the objects in our sample, RDCS0848+4453 and CDFS-CL1, have
temperatures and luminosities near the group/cluster boundary.  As
discussed in the introduction, low-temperature clusters and groups
have much lower binding masses and so should show greater evidence of
non-gravitational sources of heating.  Therefore, to rigorously test
pre-heating models of ICM formation, it will require more objects at
these temperatures and luminosities.  Surveys such as {\tt Chandra}
Multi-wavelength Project (ChaMP) \citep{wilkes2000} and the proposed
XMM survey of \citet{romer2001} will provide the necessary follow-up
of deep X-ray pointings to find the low-luminosity, low-X-ray
temperature yet high-redshift clusters and groups.  In Figure
\ref{newcl} we show an example of one such object, the new X-ray
group, CXOU J0910.1+5419.  This is one of the two systems, detailed in
the Appendix, that we found serendipitously in our {\tt Chandra}
pointings. In four fields, roughly 0.4 square degrees, we have found
two low-temperature cluster/groups.  ChaMP and the XMM survey of
\citet{romer2001} should find on the order ten or so objects a year
with these temperatures at $z\simeq 0.5$.  Such a sample will be
ideally suited for studying the evolution of the
intracluster/intragroup medium luminosity-temperature relation for low
mass systems, the same systems for which pre-heating models predict
the strongest effects.

\section{Summary}

We report the discovery and follow-up observations for \f\ and \s, two
clusters of galaxies found in the {\tt ROSAT} Deep Cluster Survey
(RDCS).  \f , which we find at $z=0.805$, appears to have a highly
elliptical, non-axisymmetric X-ray surface brightness profile based on
our {\tt Chandra} images.  \f\ is poorly described by a $\beta$ model,
implying that either we are observing a newly forming cluster, or a
recent merger event.  Our best fitting temperature is
$3.7^{+1.5}_{-0.9}$ keV with a bolometric luminosity of
$8.2^{+1.7}_{-1.6}\times 10^{43}\ {\rm erg\ s^{-1}}$.  \s , at a
$z=0.804$, appears asymmetric as well, but has the conventional
$\beta$ model parameters of $\beta = 0.49 \pm 0.06$ and $r_{core} =
165 \pm 5$ kpc.  We estimate a temperature $4.9^{+1.6}_{-0.9}$ keV and
a bolometric luminosity of $4.1^{+0.5}_{-0.4}\times 10^{44}\ {\rm erg\
s^{-1}}$ for \s .

\s\ has a strong weak lensing signal.  Our weak lensing mass estimate
is twice that of the mass we derive from the X-ray temperature and
surface brightness profile.  We speculate that we are observing a
merging cluster.  A merger along the line of sight would have a larger
total mass in the weak lensing results than implied by the assumption
of hydrostatic equilibrium for the X-ray emitting gas.  Such an event
could also serve to heat up the gas, breaking hydrostatic equilibrium.
Thus, we could be underestimating the true slope of the gas profile,
and therefore, the depth, of the potential well.
 
We combine the two temperature and luminosity measurements in this
paper with other high-redshift cluster X-ray temperature and
luminosities from the literature and create a sample with $z_{median}
= 0.83$.  We find with this sample a luminosity-temperature relation
of the form $L_x = L_6 \left( \frac{T}{6 {\rm keV}} \right) ^{\alpha}$
where $L_6 = 8.7 \pm 0.9 \times 10^{44}\ {\rm erg\ s^{-1}}$ and
$\alpha = 2.9 \pm 0.4$.  Our results agree with the low-redshift
luminosity-temperature relations
\citep{allen98,markevitch98,arnaud1999}.  If we parameterize the
evolution in the relation as $L_{x,bol}\ {\rm erg\ s^{-1}} = L_6\
(T_{x}/6\ {\rm kev})^{\alpha}\ (1+z)^A$, we find $A = 0.3-0.4 \pm 0.2$
depending on which low-redshift sample we compare against.  Such
evolution agrees with other measurements, such as $A = 0.6 \pm 0.4$ by
\citet{fairley2000} and $A = 0.0 \pm 0.7$ by \citet{mushotzky97}.  The
agreement between all of these measurements provide convincing
evidence that there is little or no evolution in the
luminosity-temperature relation. This provides more weight for ICM
evolution models that require significant additional entropy beyond
gravitational heating.

\acknowledgements

We thank Maxim Markevitch for assistance with planning our {\tt
Chandra} observation.  We thank Dan Stern for providing us with the
LRIS spectral reduction routines.  We would like to thank the
anonymous referee for useful comments. Support for SAS came from
NASA/LTSA grant NAG5-8430 and for BH from NASA/Chandra GO0-1082A.
Both BH and SAS are supported by the Institute for Geophysics and
Planetary Physics (operated under the auspices of the US Department of
Energy by the University of California Lawrence Livermore National
Laboratory under contract W-7405-Eng-48). Support for GKS in this work
was provided by NASA through Hubble Fellowship Grant
No. HF-01114.01-98A from the Space Telescope Science Institute, which
is operated by the Association of Universities for Research in
Astronomy, Incorporated, under NASA Contract NAS5-26555.  Some of the
data presented herein were obtained at the W.M. Keck Observatory,
which is operated as a scientific partnership among the California
Institute of Technology, the University of California and the National
Aeronautics and Space Administration.  The Observatory was made
possible by the generous financial support of the W.M. Keck
Foundation.  The authors wish to extend special thanks to those of
Hawaiian ancestry on whose sacred mountain we are privileged to be
guests.  Without their generous hospitality, many of the observations
presented herein would not have been possible.

\appendix

\section{The Discovery of Two X-ray Groups: CXOU J0910.1+5419 and CXOU J1316.9+2914}

We report the discovery of two new X-ray emitting groups or low-mass
clusters of galaxies.  Both systems, CXOU J0910.1+5419 and CXOU
J1316.9+2914, were found as extended X-ray sources in our Chandra
images.  Though we mention these in the text of the paper, here we
will detail their properties.

In our observation of \f , we also found CXOU J1316.9+2914, a
low-redshift, group of galaxies visible on the Digitized Sky Survey
found at $\alpha$ = 13${\rm h}$ 16${\rm m}$ 54\fs 2 $\delta$ =
29\arcdeg 14\arcmin 15\farcs 1 (J2000).  We extracted a nearby
background region and use the same model for the background as for \f
.  We found a net of 174.5 events inside the aperture for CXOU
J1316.9+2914 to which we fit a Raymond-Smith model.  The group has a
$z=0.42^{+0.14}_{-0.10}$, $T_x = 2.9^{+3.1}_{-1.2}$ keV with a flux of
$6.7^{+1.0}_{-1.0} \times 10^{-15}\ {\rm ergs\ cm^{-2}\ s^{-1}}$
(0.5-2.0 keV).  Given the large errors on the redshift it is certainly
consistent with the gas being associated with the group of galaxies
and not with a higher redshift system along the line of sight.

\begin{figure}
\includegraphics[height=6.0in]{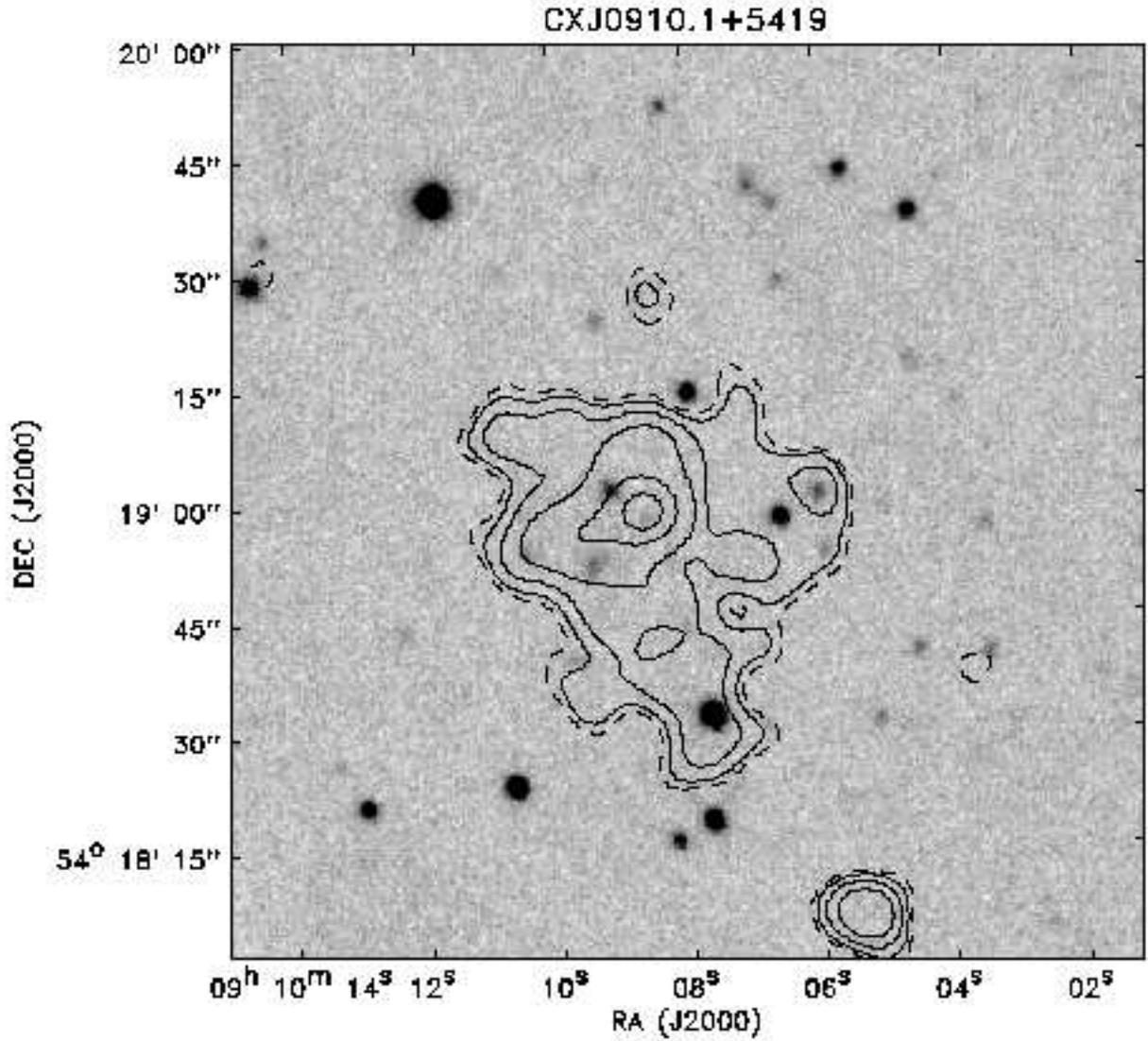}
\caption[holden.fig13.ps]{An image of CXOU J0910.1+5419 taken with the
Simultaneous Quad IR Imaging Device on the KPNO Mayall 4 meter
telescope.  The data are both the $J$ and $K_s$ registered and summed.
Plotted with contours is the 0.5-2.0 keV X-ray data smoothed with a
3\arcsec\ Gaussian.  The lowest contour level represents a $4\sigma$
variation in the background counts. 
\label{newcl}}
\end{figure}

We also found CXOU J0910.1+5419 in examining the X-ray data in
\citet{stanford2001}.  This object is interesting in that it possesses
a hard X-ray core, possibly the result of an active galactic nuclei at
the cluster center but with no apparent counterpart.  In Figure
\ref{newcl} we show the registered infrared data, obtained at the
Mayall 4m using the SQIID, with the X-ray data as contours.  We
registered the infrared and X-ray data separately with the Digitized
Sky Survey\footnote {The Digitized Sky Survey was produced at the
Space Telescope Science Institute under U.S.  Government grant NAG
W-2166. The images of these surveys are based on photographic data
obtained using the Oschin Schmidt Telescope on Palomar Mountain and
the UK Schmidt Telescope. The plates were processed into the present
compressed digital form with the permission of these institutions.}.
We found a small offset of 1\arcsec\ in both right ascension and in
declination between the X-ray data and the Digitized Sky Survey data,
the same that was found for RDCS0910+5422 in \citet{stanford2001}.
Within 30\arcsec of the center, $\alpha$ = 09${\rm h}$ 10${\rm m}$
08\fs 6 $\delta$ = 54\arcdeg 18\arcmin 56\farcs 3 (J2000), with the
central 5\arcsec\ removed, we found 464 events over 0.5-6.0 keV with
196.6 expected from the background in the same energy range.  We used
the same functional form for the background model as used in
\citet{stanford2001}.  The fit to the background yielded the same
parameters.  We fit a Raymond-Smith model with a metal abundance fixed
at 0.3 solar to those events with the background model fixed.  The
most likely result we found was $z= 0.68 \pm 0.06$ and $T_x = 2.0 \pm
0.5$ keV.  The maximum likelihood distribution is almost bimodal with
a second peak with $z= 1.18^{+0.08}_{-0.07} $ and $T_x =
2.5^{+0.6}_{-0.5}$ keV.  These are the 68\% confidence limits as
determined from 1000 Monte Carlo simulations for each set of input
values.  If we look at the overall distribution of likelihood values,
however, we see that there is large degeneracy between the redshift
and temperature for this object.  Thus, the 95\% confidence limits
encompass both combinations of redshift and temperature and all values
in between.  The observed infrared colors are more consistent with the
lower redshift, however, so we will chose that value as the most
likely.  The flux inside the aperture was $7.3^{+0.6}_{-0.6} \times
10^{-15}\ {\rm ergs\ cm^{-2}\ s^{-1}}$ (0.5-2.0 keV) and
$9.6^{+1.0}_{-1.0} \times 10^{-15}\ {\rm ergs\ cm^{-2}\ s^{-1}}$
(0.5-6.0 keV) for both temperatures, making the object almost as
bright as \f\ in terms of flux, but slightly harder.

\begin{figure}
\includegraphics[height=5.0in]{holden.fig5.ps}
\caption[holden.fig5.ps]{ Our best fitting spectrum for \f\ with the
background model subtracted and the data grouped into bins of 10
events each.
\label{rxj1317_xray}}
\end{figure}

\tablenum{1}
\begin{deluxetable}{llllr}
\tablecaption{ Results from $z>0.7$ Clusters of Galaxies\label{cl_sum}}
\tablehead{
\colhead{Name} & \colhead{z} & \colhead{Luminosity} & \colhead{kT} &
\colhead{Refs.} \\ 
\colhead{} & \colhead{} & \colhead{(bol., $10^{44}$ erg s$^{-1}$)}
& \colhead{(keV)} & \colhead{} \\ }
\startdata
RDCS0848+4453 & 1.273 & \phn$0.64^{+0.25}_{-0.16}$ &
\phn$1.6^{+0.8}_{-0.6}$ & 1 \\
RDCS0849+4452 & 1.261 & \phn$3.3^{+0.7}_{-0.5}$ &
\phn$5.8^{+2.6}_{-1.7}$ & 1 \\  
RDCS0910+5422 & 1.101 & \phn$2.5^{+0.3}_{-0.3}$ &
\phn$7.2^{+2.4}_{-1.2}$ & 2 \\
RDCS1317+2911 & 0.805 & \phn$0.82^{+0.17}_{-0.16}$ &
\phn$3.7^{+1.5}_{-0.9}$ \\
RDCS1350+6007 & 0.804 & \phn$4.1^{+0.5}_{-0.4}$ &
\phn$4.9^{+1.3}_{-0.9}$ \\
CDFS-CL1 & 0.73 & \phn$0.13^{+0.05}_{-0.05}$ & \phn$1.2^{+0.6}_{-0.3}$
& 3 \\
MS1137.5+6625 & 0.782 & $12.6^{+1.5}_{-1.5}$ &
\phn$5.7^{+0.8}_{-0.6}$ & 4,5,6 \\
RX J1716+6708 & 0.813 & $10.3^{+0.5}_{-0.5}$ &
\phn$5.7^{+2.5}_{-1.5}$ & 7 \\ 
MS1054.4+0321 & 0.833 & $34.3^{+1.9}_{-2.1}$ &
$10.4^{+1.3}_{-1.2}$ & 8 \\  
RDCS0152.7-1357 & 0.831 & $13.0^{+0.3}_{-0.3}$ &
\phn$6.5^{+1.7}_{-1.2}$ & 9 \\  
1WGA J1226.9+3332 & 0.89 & $29.3^{+3.0}_{-3.0}$ & $10.5^{+4.0}_{-3.0}$
& 10,11 \\  
RX J1053.7+5753 & 1.26 & \phn$3.4^{+0.34}_{-0.34}$ & \phn$4.9^{+1.3}_{-0.9}$
& 12 \\  

\enddata

\tablerefs{(1) \citet{stanford2000}; (2) \citet{stanford2001};
(3) \citet{giacconi2001}; (4) \citet{borgani2001}; (5)
\citet{gioia94}; (6) \citet{donahue99b}; (7) \citet{gioia1999};
(8) \citet{jeltema2001}; (9) \citet{dellaCeca00}; (10) \citet{cagnoni2001};
(11) \citet{ebeling2001}; (12) \citet{hashimoto2002}}

\end{deluxetable}

\end{document}